# Expounding on physics.

# A phenomenographic study of physicists talking of their physics

*Åke Ingerman*
*Department of Microelectronics and Nanophysics*
*Chalmers University of Technology*

*Shirley Booth*
*Centre for Educational Development*
*Chalmers University of Technology*

## Abstract

Physicists and physics students have been studied with respect to the variation in ways they expound on their topic of research and a physics problem, respectively. A phenomenographic approach has been employed; six fourth-year physics students and ten teacher/researcher physicists at various stages of their careers have been interviewed. Four qualitatively distinct ways of expounding on physics have been identified, constituting an outcome space where there is a successive shift towards coherent structure and multiple referent domains. The interviewed person is characterised as expressing an 'object of knowledge' and the interviewer is characterised as a willing and active listener who is trying to make sense of it, constituting a 'knowledge object' out of the ideas, data and personal experience. Pedagogical situations of analogous character to the interviewer-interviewee discussions are considered in the light of the analysis, focusing on the affordances for learning offered by the different forms of exposition.

## Introduction

Scientists and science students are constantly required to tell others about physics, whether it is the physics they are researching, the physics they are teaching, or the physics they are learning. Experience tells us that some do a better job at it than others, and that one day a better job might be done than on another day, and that a better job might be done on one theme than on another. There is a variation in the forms of explanation or, more precisely, exposition[1], that occur across and within the collective of our physics colleagues and students of physics. This paper attempts to bring some clarity and analysis to the issue by reporting a study of the ways in which physicists[2] and senior physics university students expound on physics problems and physics research.

The work reported here is a part of a larger study, the overriding research question being, 'What does it mean to become a physicist?' In this particular report we are addressing a subsidiary question: 'In what ways do physicists experience physics problems and research?', and the more specific question 'What forms of exposition do physicists use in discussing their physics problems?'

---

[1] We prefer 'expound' and 'exposition' to the more obvious 'explain' and 'explanation' for two reasons. First, they are more precise, and second they avoid confusion with the 'forms of explanation' associated with grand theories.

[2] In this article, we take physicist to mean an active physics researcher.

The question has a pedagogical relevance in that one major form of physics education comprises expositions of physics phenomena, whether teachers are expounding for undergraduates, researchers are expounding for their doctoral students or colleagues, or students are expounding for their teachers in an examination setting. It has a physics relevance in that physicists are daily engaged in efforts to convey the meaning of the physics they are working with or teaching, and we hope to offer tools for the analysis of what sense-making is thus afforded.

The study was carried out with a predominantly phenomenographic approach (Marton, 1981; Marton & Booth, 1997; Booth & Ingerman, in press). This implies that the object of research is the variation in the ways certain phenomena are experienced or handled by people of interest. In this paper, the group of people of interest comprises physicists, in particular physics students towards the end of their education and physics researchers at various levels of experience. The phenomenon is advanced problems in physics (in the case of students) and current physics research (in the case of researchers), but the aim of the study is not to analyse and describe the ways in which the group experience what constitutes the phenomenon, but rather how they handle the exposition for a discussion partner.

## *Method*

The data that has been analysed comes from interviews that were conducted with students and researchers at the Physics department at Chalmers University of Technology and Göteborg University, Sweden by the first author of this paper. The interviews were held with 6 fourth year physics university students and with 10 physics researchers in the sub-field mesoscopic physics of condensed matter physics, chosen to represent a variation in research experience, from doctoral students to professors, both theoretical and experimental.

The interview as a whole was designed to be semi-structured around a number of predetermined themes and open to exploring these themes in ways that might emerge in the particular interview. In line with the phenomenographic approach, the interviews are seen as forming a 'pool of meaning' in which the variation in ways of experiencing the phenomena of interest for the group, and at the collective level rather than at an individual level, are to be seen. The total interviews varied in length between 45 minutes and two hours, mainly depending on the length of time the problem or research, which was the central element of the interview, was discussed. They were recorded on audio tape and on video tape, so that the jottings, the sketches and some of the movements of the interviewees were captured. The interviews were conducted in English, largely because much of the advanced education is in English and it is the natural language for discussion; in addition, the practical issue of needing to translate the interviews into English if held in Swedish played a role in the decision. The interviewees were actively encouraged to speak in Swedish if they felt it would help them express themselves on a particular point. All of the interviewees were comfortable with this, though some of the students hesitated more than they might have done otherwise.

The whole of the study includes several analyses, addressing different issues, the two main ones being about the form of exposition (reported in this article) and how physics research results are established among fellow physicists (reported in Ingerman, 2002). The element of the interview that is of specific interest for this paper was different for the group of students and the group of researchers. The students were given a particular abstract problem that they were asked to talk about in several ways (described more fully in the next section), while the researchers were asked to describe their current research topics, which naturally varied within the group.

In general, in phenomenographic analysis of text-based data, we proceed by reading the interviews repeatedly with a particular aspect of the interview theme in focus, now as expressions of individual students, now as series of extracts related to specific issues. Ways of understanding them evolve, categories are formed and reformed, extracts from interviews are sought to support and give substance to the categories, and logical and empirical links between categories are explored. Finally, an outcome space is presented in which the categories are described, the structure indicated by logical links is analysed, and the empirical links with the data and the research situation are offered as illustrations and elaborations of the meaning of the categories.

The aim of the analysis presented here is to offer an outcome space which captures the essence of the exposition of physics and reveals the essential variational structure of that experience at the collective level. By 'collective level' we mean that we are not describing and categorising individual's ways of expounding physics problems or research topics, but are rather describing the forms of exposition that can be found across this group of physics students and researchers. It was not the aim of the analysis to produce general and incontrovertible results, but rather results that are both valid and interesting in the local setting, lending meaning to the specific situation being examined, as well as a contribution to the more general understanding of pedagogical issues related to teaching and researching physics at the university.

Focus is, in the first place, on the structure of the expositions, abstracted away from the immediate content, thus making expositions of different kinds of physical situations comparable, but nevertheless retaining the physics referent[3] and content. The data was analysed by first considering extracts from the interviews and noting significant characteristics of structure and meaning. The variation thus revealed was refined by relating back to the data once more and looking explicitly at the structure, and at the referents of the structural components. These structural components can be characterised as sequences of discussion where we can identify a 'type of talk' related to the research topic or problem, and also see its boundaries, where the discussion goes off to another 'type of talk'. The different 'types of talk', and the ways in which the referents are related to the object being discussed, thus came to form the basis for categorisation. Thus a system of categories of description of qualitatively distinct ways of expounding on the physics problem or research was arrived at, an outcome space.

During the analysis of forms of exposition, the interviews naturally fall into two subgroups, those of students and those of researchers, in that, as already mentioned, they were asked to discuss distinctly different features of their physics experience. Nevertheless, in the analysis strong similarities were found in the variation of the qualitative structure of the expositions in the two subgroups. First we will describe categories of description for the student group, then categories of description for the researcher group, and then consider them as a single outcome space. In the present article, quotes from the interviewees are included to illustrate the essence of the categories of description, and can not be regarded as a full material for analysis.

## *The study of students discussing a problem*

The students, who were all at the end of their fourth year of university studies, were engaged in a discussion around a problem which they had inevitably encountered during their third year course in quantum physics. It can be called 'the barrier problem'. It was introduced, typically by the interviewer (I) asking, as here, taken from the transcript of an interview:

> I   I would like now to ask a number of questions around a certain physics problem. Say that I have an object of some sort, it is moving here (figure 1), and this object, it feels some kind of energy landscape or potential landscape, and here is some kind of obstacle which is in the path of this object. And the question is, how is the object affected by, the motion of the object affected by, this obstacle? This is the question, but this is, I would say, it's quite an abstract way of formulating the problem and I would like to connect to some physics. Do you recognise this kind of problem?

---

[3] By 'referent' is meant that to which the extract of discussion refers.

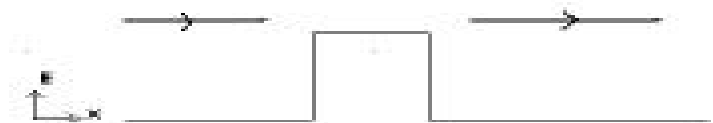

Figure 1. The sketch of 'the barrier problem' given to the students

The students had met this problem as a textbook problem in a course in quantum physics in the third year, and all of them recognised it from this context. However, the problem can also be relevant in situations outside quantum mechanics. The problem is formulated in momentum space rather than in coordinate space, a fact that was not recognised by most students.

Posed in this way, the presented physics problem does not refer directly to any specific situation, and this part of the interview opened by asking the students to give examples of physical situations where the model might be relevant. Then they were asked to discuss, in the case of the quantum physics situation, the probability of transmission as a function of energy of the object, which in this case was identified with an electron, approaching the obstacle.

They were also asked, specifically, to write down a mathematical model for calculating the transmission probability as a function of the energy of the electron, and to indicate how to it could be solved. The students embarked on the standard way of approaching the problem, but were not, quite naturally, able to solve it fully in the interview situation, without having any textbook to rely on. The interviewer delivered the result, given by the textbook the students had had in their course on quantum physics, Eq. (1).

The standard way to solve the problem is to use the Schrödinger equation to describe the electron. Using an ansatz of plane waves as the electron wave function, the transmission probability can be found to be

$$T=|\psi_{in}|^2/|\psi_{out}|^2=1/(1+ (k_1^2-k_2^2)^2\sin^2(k_2 L)/4k_1^2 k_2^2) \quad (1)$$

where $k_1$ and $k_2$ are the wave vectors related to the momentum of the electron in different regions and L is the width of the barrier. The wave vector of an electron is related to its relative energy E (compared to the reference energy which is indicated in Figure 1 with the thin 'ground' line), and is different outside the barrier, where $k_1=\sqrt{2mE/\hbar}$, and inside the barrier, where $k_2=\sqrt{2m(E-V_0)}$, where m is the mass of the electron, $\hbar$ is Planck's constant, and $V_0$ is the height of the potential energy barrier. It was pointed out to the students that Eq. (1) was written in compact form, and valid for all energies.

Considering the problem in classical physics, the probability of transmission is either unity, if the energy of the object is larger than the height of the barrier, or zero otherwise. In quantum physics, the probability can be less than one even when the energy of the electron is higher than the barrier, and larger than zero when the energy is smaller, which is called tunnelling. The probability of tunnelling decays exponentially as a function of the height of the barrier and as a function of the width of the barrier.

Analysis of the interviews with students revealed an outcome space of four categories of description which can be logically and hierarchically related to each other, as we will describe later. They are provisionally labelled Categories A, B, C and D, with subscript S for Student, but will be renamed in due course to indicate their main characteristic. We want to emphasise even here, that these are not categorisations of individuals, but categorisations of ways of expounding physics problems and topics, several of which can be voiced by a single interviewee in different contexts and even by the interviewer who is also a participant in the discussion.

## *Outcome space for students*

### *Category $A_S$*

In this category, explanations of the problem are given, without anchoring and internal relation. The structure is chopped and the referents of the structural components are unrelated. Bits of the context of the problem are picked up and referred to in a sense of name-dropping.

Here a student is introduced to the barrier problem, as described above, and responds:

> $P_8$ …[4] Yeah, from quantum physics, where we can have a barrier for instance, you have an electron, electron around the nucleus.

Here two different physical situations are mixed: 'quantum physics', 'a barrier' and 'an electron around the nucleus' are named as elements that occur in the stated question and the diagram, but without being elaborated or related to one another.

> I  Could you elaborate a little bit on what kind of physical situation that could be relevant for this description? Where it could be interesting to look at this … problem?
>
> $P_8$ So, this particle is of course affected yeah it's [inaudible][5], some and ... yeah for the electron and the nucleus maybe it's not a barrier like this for the electron to come free. … So it's a particle here moving, do you see it as a motion like this and then it's coming into the barrier?

When the interviewer tries to prompt for a connection to a particular physical situation, the student tries to sort this out, with various comments, before realising he[6] does not know what he is explaining about and asks the interviewer to clarify the situation.

---

[4] … in an interview extract indicates a lengthy pause; shorter pauses have been indicated by punctuation

[5] Comments in square brackets have been added by the authors for clarification

[6] Most of the interviewees were male, and so where the use of a personal pronoun to refer to a specific person is unavoidable, the masculine is used to avoid exposing the female interviewees. The interviewer (Ingerman) is always referred to by a masculine pronoun.

I   Yeah, I see it as the object has some kind of movement some kind of … speed which is related to its energy, so it's moving towards this obstacle, or barrier.

$P_8$   I can't think about any situation right now.

That intervention from the interviewer was not of much help, and he continues:

I   Is there somewhere that it could be relevant outside quantum physics, for example, just to widen the possible physical situations?

$P_8$   Yeah, … I'm thinking about electronics or if it, if you have a current and in, I haven't really thought about this before like this, but if you have a current and then if you need to go maybe through a diode backwards and then you need a really high voltage to come through the diode. The diode is like a barrier.

This is a case of hopping from one bit of the problem to another in potential ways of discussing or explaining the problem, without finding a coherent story to offer. This lengthy example makes a substantial contribution to constituting this category, and there are only glimpses of it elsewhere.

### Category $B_S$

In this category there are lengthy components to the exposition which maintain a single perspective, and there is a sharp discontinuity between components of the exposition. The referent of a component might change, but always identifiably within one domain, either physics or mathematics. One frequently occurring example is that of carrying out a calculation, which is anchored in a mathematics and is carried on without explicit reference to the physics to which it refers.

A mathematically focused manifestation refers, explicitly or implicitly, to mathematical concepts or objects or to formal calculations, with phrases like 'is a Fourier transform' or 'behaves like a sine', focusing on the mathematical formulation and description, as in the following example. Here a student is asked to evaluate the resulting formula for the transmission in the 'barrier problem', Eq. (1), what the implications are, and if the results are reasonable:

I   Can you see that the transmission is less than 1, if you are not, E is smaller than $V_0$. Is it consistent?

$P_{11}$   It should be. I have to think. I mean it is ... yes it is. You see, this part here will always be greater than 1 than 0 I mean, it will always be positive, now if this is, if E is less that $V_0$ this [$k_2$] would be imaginary but that means that sine [sin $k_2L$] square will also be imaginary so the imaginary parts take each other out and I have something real, which is also greater than zero, because this is always greater than zero, and this will never be imaginary, times something negative, it will be something positive still, so it is always also

I   And sine is never equal to zero?

$P_{11}$   Sine is equal to zero only when this [$k_2L=n$  ] is fulfilled.

I   Which is not true.

$P_{11}$   That's right.

I   OK that's consistent.

This is pure mathematics all the way through. What the formula describes is taken for granted, by both the interviewee and the interviewer.

An exposition in this category with a physics focus, in contrast, refers to physics concepts[7], like energy or momentum, and takes its strength from physics phenomena, referring in some way to the world, not necessarily tangible or visible, possibly even very abstract, but nonetheless recognisably physics rather than mathematics. The following is an example of an exposition referring solely to physics concepts:

>    I   I, tell me how you would design an experiment. What is it that you would try to observe, or?
>
>    $P_6$   When trying to verify the model as such? Well, I mean, the principle that lies behind a sweep-tunnelling microscope for example, I would say is a quite a good way of actually proving that this works to some extent at least. I mean, having an extremely thin tip of some kind with a high charge and moving it across, along some sort of surface at a very, very, small distance and, but still a distance, so that this vacuum thing between it would, I mean, would be some sort of barrier like this, and still observe that there are passing electrons from the tip down to the surface. That is, I mean that's, I don't know if it would call it an experiment, it's an application, but it's still...

In both of these examples the interviewee has given rather a lengthy response that focused solely on either the physics or the mathematics, without bridging between the two. Even if the speaker is aware of the problem from many perspectives, he or she is only expressing one of them.

### Category $C_S$

In this category the exposition maintains a coherent and continuous structure while the referent can shift between different domains, in this case the domains of physics and mathematics. It has none of the choppiness of category $A_S$, nor the long and mono-perspective of category $B_S$.

Here, a student is evaluating the transmission formula, Eq. (1):

>    I   and could you add something more from looking at this solution?
>
>    $P_7$   […] So the transmissivity ...first, the first test we can do is if $k_1$ is equal to $k_2$, this would be one, what happens then it just pass and that's the same as V would be 0 so that's OK. ...Depends on the sine here so it could have some kind of oscillating behaviour depending on the width of the hill. I have to think about this...$k_2$ is real when this and when the energy of the electron is higher than the potential of the hill, so it's just an oscillating behaviour if the energy of the electron is higher than the hill, so then it's supposed to pass and the wave it will be, the rest of the behaviour must depend on the wave behaviour of the electron that depends on the width of the hill.

During his evaluation of the transmission formula, the student connects the behaviour of the transmission in different cases to the physical meanings of the parameters, e.g. the width of the hill. This indicates a simultaneous awareness of both the physics of the problem and the mathematics of the formalisation, and conveys both to the listener. The behaviour (oscillation as a function of the width) surprises him, and he does not seem to be able to give an explanation of why this happens.

---

[7] As physics concept we also count concepts which are anchored in both physics and mathematics, like a wave function (if not used in a specific mathematics way).

Here is an example where the transmission probability formula is being evaluated in physics terms, and references are also being made to the mathematical side of it:

> I   But you mentioned earlier that there was some values here that it would actually be one.
>
> $P_6$   Mmm, but that would then, it would coincide with multiples of the wavelength and for these particles a whole wavelength is apparently a very nice property. But that, it would be one for those, for example those that had just one path or what ever you say. But also the multiples.
>
> I   You said that when the energy is smaller than the barrier, then it's different.
>
> $P_6$   Yeah, then we will end up with this term, being imaginary, (OK)[8] and then would be, then the sine square actually would be a, some sort of exponentially declining function.

### Category $D_S$

In this category an exposition extends and/or evaluates a situation and/or the posed problem, going beyond the given situation in some sense; it might even point to a reformulation of the model and explore new dimensions. Thus the exposition comes from outside the problem at hand, and the problem, or some aspect of the problem, is related to some new context. The structure is coherent, as in $C_S$, and the referent is the problem or an aspect of it.

This example is from a comment on the solution of the barrier problem, where both the interviewer and the interviewee are engaged in bringing forth such an exposition:

> I   OK. You also in the beginning mentioned here about the time dependence of this problem that you should actually it's realistically not to get plane waves but rather wave package coming in. Do you have any feeling for, what kind of difference would it make to have a wave package instead of a plane wave, and how realistic is a plane wave when you...

The interviewer thus initiates an extension of the solution of the posed problem, and discusses its relevance, taking a perspective on what it means to solve this particular problem as well as this kind of problem in general. The interviewee responds:

> $P_9$   Yeah, What you mean by wave package is for one thing is that if you think in the terms of Fourier analysis that a wave package is just a superposition of plane waves with different momentum and which can give some kind of structure generally of the look at anything. But what one usually looks at is the Gaussian wave package, which is kind of, behaves more like a classical particle, in the sense that for example, I mean, as far as nothing happens so to say, even this wave package is moving, though it is spread out it will not, the momenta here will be quite well defined and so on and also the [position]. And you can see that that part of the wave packet has that momenta and so on and, but when there is scattering, then what we see here from that part of the transmission is that everything is k-dependent and this one is made up from different k's and that will lead to that the wave package is kind of, I mean it will lead to this kind of dispersion of this wave package, because different k's will, the plane waves that correspond to different k's have different transmission coefficients in a sense, at least for this simple kind of barriers, what will be transmitted will be a wave package on the other side and in fact there will also be a wave package going in the other direction.

---

[8] Text in round brackets in the interview extracts is a comment from the other party.

He continues to discuss what might happen in different situations, using the calculations discussed as a starting point.

This form of explanation was only rarely seen in the students' interviews, and in fact the quoted subject was a new research student, categorised as a student because of the interview form he took part in.

## *The study of researchers discussing their research*

The interviews that were held with researchers were intended to let them expound on the salient aspects of their current work. The interviewer thus allowed them to choose freely, in contrast to the students who were given a fixed problem. However, they were all in mesoscopic physics – a sub-field of condensed matter physics – either in the interviewer's, or a neighbouring, theoretically oriented group, or in neighbouring experimental groups.

### *Outcome space for researchers*

The researchers' expositions of their current research was analysed with the same focus as the students' expositions of the barrier problem. Even though not all of the researchers discussed the same research problem, and though they stood in a different relation to their problem than the students to the barrier problem (they, and not the interviewer, introduced their topic), the forms of exposition could be analysed into similar categories, except that category A was, not surprisingly, not seen at all.

#### *Category $B_R$*

This category is, again, characterised by structure – a lengthy, and sharply delimited, maintenance of focus on one aspect of a problem. In the students, manifestation of a mathematical and a physical nature were seen; here in the researchers' expositions these are also to be found, with an additional technical or experimental referent, where constructing something in the real world, or a scheme for getting something to work is taken up at length.

A researcher, expounding what his research is about, refers solely to physics concepts in the following extract

    I    Could you please just sketch the device for me? (OK) Just to make sure that I know what it is.

    $P_3$    You have a single cooper pair box (mm), which basically is a small superconducting island. I could put S[uperconductor] there, then it's connected to a superconducting lead reservoir, which you make large and which you make ground or something and then you have, oops, capacitance, gate capacitance and then you have some gate voltage, there and then you have some, you have the Josephson coupling between these two, you are familiar with that? (mm) And then you have some, this island is so small that you have the charging energy in the qubit which is…, so the charging energy of [the qubit] should be large (mm) so if you plot the energy here versus gate voltage…

Here a series of physics concepts, e.g. single Cooper pair box, superconducting island, superconductor, reservoir, and their associated properties are linked and used in expounding what the researcher's research is about. He continues for a while, describing the situation with a physical character of the exposition.

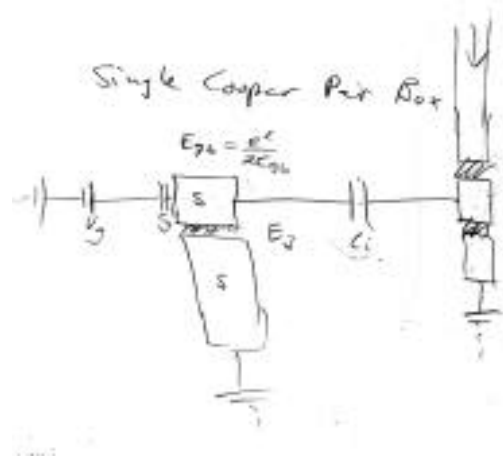

Figure 2. Sketch of a single Cooper pair box produced by $P_3$, referred to in category $B_R$

The following extract is from a researcher who is expounding on what he is calculating, and concentrates for a long while only on mathematical aspects of his research problem:

$P_5$  Yes there is, I mean the total Hamiltonian of the system is Hamiltonian for sub system one [$H_1$], Hamiltonian for sub system two [$H_2$] and the tunnelling Hamiltonian [$H_T$], so these guys that appear here, their dynamics is coupled to the rest of the system, with these Hamiltonians that we describe the independent subsystems.

I  And the time development of $H_T$ is determined by $H_1$ and $H_2$ and the exponential, OK, then I see how, the problem crops up. OK and how far have you come in your calculations, you said that you integrated out both the boson operators?

$P_5$  How far I have gotten in the calculations is that I have expressed this current in terms of a few quantities one of them is what I call the spectral function of tube one and one of what I call spectral function of subsystem two and one of them is the or two others are the density-density correlation function so these guys are just minus two times the imaginary part of the retarded Green's function of tube one at some frequency and this guy is minus two times, OK, let me say the imaginary time [inaudible] is a density-density correlation function. So from here you need take the retarded one. So I have an expression now where this is expressed in terms of the simple quantities…

This researcher explains his research in terms of a technical character:

I  Now I would like you to tell me a little bit about your area of research.

$P_{16}$  Yeah. I'm working with small Josephson junctions. First of all I work with the RF-SET, a single electron transistor and what it basically is a SET transistor, is transistor structure in principle that's having tunnelling with single electrons. That's why it gets its name. And it's a field transistor, a field effect transistor in principle and this one has, to have it work we need to cool it down so we work in cryogenics so we work down at 20 millikelvin or at that range, the millikelvin range. And this means that there are a lot of additional problems to keep the sample cool, and to get down leads to actually do measurements on our samples and that turns up to be an electrical problem also because attenuation and filter effects in your leads going down, it's almost 2 meters of wires going down, and 2 meters going up.

What is discussed in the above quote is a scheme or the way in which a specific device is to be made to work. What is the meaning and reason to get it to work is taken for granted.

### Category $C_R$

This category resembles that of the corresponding category for students, in that there is a multiple focus in the form of exposition. But whereas the students, constrained by the problem they were dealing with, brought physics and mathematics together, the researchers additionally juxtaposed physical and technical perspectives on their research.

This example is from the researcher quoted above in category $B_R$, who when continuing to tell about his research relates to both physical and mathematical aspects, sketching at the same time figure 3.

I  Let me, since you started one pure state?

$P_3$  You start in one pure state and then you quickly turn your gate voltage, so that this pure state here of zero extra, will project, here you have, here your basis set is different.

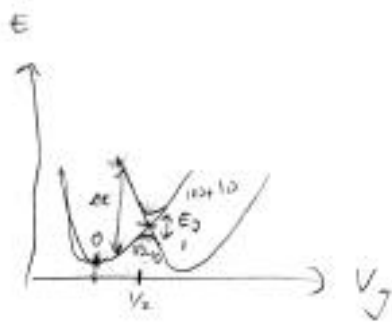

Figure 3. Sketch of energy levels produced by $P_3$ and referred to in category $C_R$

Here something new occurs. Relating the physical state to the mathematical description in terms of a basis set, makes it possible to connect the picture of occupying a state to the possibility of projecting this onto non-diagonal states. Thus a mathematical understanding and a physical understanding of the same situation is linked and integrated.

Here is an example of an exposition where technical and physical arguments referents are juxtaposed:

I  Just, this is the island, the box and this is coupled to ground through controllable tunnel junctions? (Mm.) OK. And this box is connected to the readout system? (Yeah) OK, so, and this readout system is also a super conducting island? (Yes) Which is coupled to some frequency generator or readout, so you send in a frequency and then you get some kind of current response or what kind of response do you get? How does the readout work? I'm not really good at electronics.

- P₁₃ Well, the readout system also of course includes a gate for the SET[9] here. I mean with this gate you can tune the SET. You can tune it to be in an open state or you can tune it to be in the closed Coulomb blockaded state. (OK) And if the set is in the blockaded state there is no current going through the island this way, so everything is reflected back of the microwaves, but if you open the transistor slightly with the gate you will get some dissipation down to ground and less signal will be reflected, but... So if you just hold the gate voltage constant, if this, if the charge on this island changes that will electrostatically affect the island...

Interesting here is to note e.g. 'with this gate you can tune the set. You can tune it to be in an open state or…', which connects the tuning of the device, i.e. to turn knobs, with the physical effects of that, to put the system into a state. How the device works is connected what physical meaning it has.

### Category $D_R$

As with the students, this category contains expressions of research where it is compared or contrasted with other research, or introduced into new contexts. In the following extract, a researcher lifts his exposition to relate it to other approaches to the same sort of problem and criticise weaknesses that can be seen there:

- I   So you want to understand the same problem, but in a different way?

- P₂  Yeah, in order to, because that way would be a better way to describe the physics mathematically, because it is on solid ground with broadening and that stuff, some sort of ground and you can really calculate stuff instead of doing it phenomenological. So in a sense it's an old problem and already solved, but on the other hand it's new and the calculation should be done at some point.

- I   What is the purpose of doing these calculations?

- P₂  Well, first it's, for me it's frustrating when people in the d-wave superconductivity business are very sloppy in tunnelling, they are not well educated in tunnelling as in mesoscopic physics. So maybe it's worthwhile to do this calculation because it's important when you have surface bound states and the mid gap state. Then it could be relevant because if you do calculations of the noise you would see that it's different zero frequency shot noise if you have broadening or not. Then broadening becomes measurable so it would be a way of probing the broadening of the superconductor. So if this broadening is due to surface roughness then it's not so interesting because, well you have measurement of surface roughness but that is not so, that is not of fundamental interest, but if the broadening has, is connected to the scattering against some boson bath or something like that, then you probably [inaudible] this bath. But this could be interesting in high $T_C$ [high temperature superconductivity] because it may be some spin fluctuation magnon bath instead of just the formula for example and that could be connected to more fundamental questions of superconductivity. So in a sense it could be useful in that way.

Here P₂ compare the physical effects when considering different models (e.g. with or without broadening). This can then in turn be measurable, and connecting back that result as input to the calculations, as a sort of evaluation. We see this as an example of going beyond taking two perspectives simultaneously in an exposition, to going outside the problem and seeing it as a whole in a wider context of related research.

---

[9] SET: single electron transistor

The structure can be said in this case to be reflective and the referent is simultaneously the problem, the situation the problem arises in, and the relation between then.

## *Discussion*

First we will unite the two outcome spaces and then we will discuss their significance for pedagogical issues.

### *The joint outcome space*

An outcome space comprising four categories of description has been identified from the data gathered from physics students and one of three categories from that of the physics researchers. These are very similar, the major differences being that the first of the students' categories does not appear in the researchers' data and the other three are more elaborated in the researchers' data. The extracts of dialogue have been categorised according to their structure in the first place and the reference made within the structure in the second place, and they have been illustrated with extensive extracts from interviews. Let us now bring the outcome spaces together, adopting a consistent terminology and considering the relation between the expounder and the listener:

*A. Expounding in bits ($A_S$)*

Here the physics problem being discussed is treated fragmentarily, the structure is fine-scaled and chopped or incoherent. Talk goes from one theme to another with unlinked referents. It was seen only rarely among the students who were given a problem situation to discuss, and not at all in the researchers. It indicated an uncertainty of how to see the problem, where to start, what perspective to take, what to bring into focus and what to leave in the background.

For the listener of the exposition, the problem remains fragmented, and probably unclear. There is no clear message to be obtained from the exposition.

*B. Expounding in a single perspective ($B_S$, $B_R$)*

Here a single perspective on the physics problem being discussed is held in focus at length, whether a physical, mathematical or technical perspective, and is sharply separated from expositions from other perspectives. Thus the structure of sequences of language in this category is large-scaled and coherent, and the referent remains consistently within one domain until the exposition changes track and then another referent domain can take over, or exposition from another category type takes over.

The listener is offered a consistent view of the problem or topic from a single perspective, at the risk of losing contact with its complexity. In particular, both the speaker and the listener can lose track of the problem as physics while delving further and further into the mathematical formalisation, and hence the sharp boundaries as one or the other pulls the drift back to the problem, or a satisfactory conclusion is reached in the single perspective and the wider discussion can be resumed.

*C. Expounding in multiple perspectives ($C_S$, $C_R$)*

Here two (or potentially more) perspectives are juxtaposed during exposition. The structure is coherent, there is none of the choppiness of the first category; the referent is dual or multiple, typically mathematics and physics or physics and technology, and there is none of the long single-focused exposition of the second category.

The listener is offered an exposition in which both, or several, perspectives are taken, and less of the complex problem is taken for granted. It is possible for the listener to see the object of discussion from more than one perspective at the same time, and this opens up a greater possibility to build a nuanced picture of the object.

*D. Expounding through contextualisation ($D_S$, $D_R$)*

Here the speaker has shifted outside the problem or research topic at hand to compare it with similar or different problems or topics, or view it in a new context. The structure is coherent and the referent is the whole object of discussion at hand as well as similar or related objects.

The listener is now offered a greater range of perspective. The contextualisation that takes place offers the listener other experiences, possibly more familiar, to relate the object to, and possibly more familiar situations from which to view the problem in hand.

### *The structure of the outcome space*

As already said, the object that was expounded on was different for the two categories of interviewee: the students discussed a problem in quantum physics and the researchers discussed their own research. Some readers might ask themselves: 'Why do some of the students appear in categories that are in some sense more advanced while the researchers appear in categories that are less advanced?'. We will address this point with two arguments.

First, we are not categorising individuals, or even the total ways in which particular individuals expressed themselves. We are categorizing extracts from expositions of the object for the interview, and in many cases, individual interviews gave rise to several forms of expression and thus contributed to more than one of the categories. In the illustrative extracts, expressions from student $P_6$ appear in both categories $B_S$ and $C_S$, and expressions from researcher $P_3$ appear in both $B_S$ and $C_S$.

Second, we might well be tapping into an underlying general structure of experience here, which both students and researchers are aligned with. Namely, if there are two inherent features to a phenomenon then there are four ways in which they can be experienced: total absence, presence of one, presence of the other, and presence of both. Now that might appear to be trivialising the results we have derived inductively from the data, but might nevertheless apply. The importance of our results lies less in the number of categories or even the structure of the categories, but rather in their descriptions and significance in the pedagogical contexts to which we apply them.

As one example of this we can point to another study, of students learning to program, where there were also four categories, this time of initial approaches to tasks of writing programs to satisfy given problems (Booth, 1992). These were labelled and described, in the context of that study, as expedient (focus on finding a solution), constructual (focus on the programming language alone), operational (focus on how the program will be executed and what programming functions are needed) and structural (focus on interpreting the problem in its own domain). This can also be simplified to the underlying structure, if we say that the experience of setting off on writing a program has two inherent features: the problem and the program. Now we have the approaches as expedient (both absent), constructual (only programming present), operational (both problem and programming present) and structural (only problem present). Now, in that pedagogical situation, namely beginners learning to program for professional purposes, it was the fourth of these that was identified as most significant, where the only one inherent feature of the experience of programming is present – the problem. In the present study, as will be revealed below, we identify the combination of both present as pedagogically most rewarding.

Here we will also point out that the categories, showing as they do an underlying structural similarity between students and researchers, generally have a richer and more complex content in the case of researchers. For them, particular expressions are constituted of more features and they are more intimately connected to one another. This is also born out by the observation that category A, which is particularly poor in content, was not seen in the researchers' expositions on their research.

### Interviews as discussion

Phenomenographic interviews are characterised as being semi-structured and open, which is to say that while there is a carefully planned scheme of openings to explore the interviewee's relationship to the phenomenon in question, the interviewer takes opportunities to follow up promising leads, and to refer back to earlier points of discussion. Thus both interviewer and interviewee are engaged in a process of making mutual (if not common) sense of the phenomenon.

That was the situation in this study, where the interviewer was engaged in following the interviewee's line of reasoning and the interviewee was trying in various way to express his or her understanding. In the researcher interviews, the interviewer was familiar with the material in outline, to varying degrees, but the specifics were hazy or unclear, and then he had to find expression for his understanding or lack of understanding. In the case of the student interviews, the interviewer was thoroughly familiar with the situation being dealt with but had to explore the student's own understanding, difficulties and interests. For example:

> I  Just to, it seems a little bit strange to me, you are introducing k as related to momentum and you say here that the sign here of the plane wave is the same for the incoming and the reflected and that is they have the same momentum, but to me it seems that something coming in and something being reflected, going the other way should have opposite signs of the momentum.

Here, it is the interviewer who connects the mathematical representation and the ongoing calculations to the physical situation – going opposite ways, momentum – in an exposition that can be placed in category C, multiple perspectives. The interviewee goes on to agree, more or less.

> $P_9$  OK, one can of course one can put something else here, k prime. That's perfectly all right but I guess will come out correctly in the end. So that's all right.

Thus, the interviews often had the character of a discussion between two interested parties, and the forms of exposition described here apply equally to interviewer and interviewee. Further it can be seen that the form employed by one of them can encourage the form of the response. Which, once again, leads us to say that it is the forms that are significant as structures, inextricably intertwined as they are with meaning, and not who says something or why they say it.

### Structure and meaning

Differentiating the *meaning* of the exposition from the *structure* is not unproblematic, and as we have said earlier, our analysis is, in the first place, according to structure. To exemplify this we can return to the description of category $B_R$ and the illustrative quote from $P_5$. This is an example of a lengthy mathematical exposition. And if we look at the continuation of that extract, we see that $P_5$ continues to expound on the calculations involved in the research problem, as indeed does the interviewer:

> I  Just to give me an idea of what you calculate actually.
>
> $P_5$  OK, let's see. There is a, if you look in Mahan's book on many body physics you can find an expression called the thermally average current as … some constant … me thinks it is 2 over h-bar [$2/\hbar$] but whatever, times a retarded imaginary part of a retarded correlation function evaluated at frequency eV over h-bar where eV is the applied frequency. This retarded correlation function in time space is minus i beta [-i_] and then there is a commutator between two terms $H_T$ of the [inaudible] and $H_T$ (OK) Hermitian conjugate of zero where $H_T$ is a tunnelling probability so we'll take let's call some subsystem one and one subsystem two and then there is some tunnelling term here and then there is the Hermitian conjugate, and something like this. Now in our case this is more complicated due to this tube bending.

But look, just at the end of this highly mathematical description there is an indication that throughout his mathematical description, even that which we have categorised as $B_R$, expounding in a single perspective, he might all the time have been aware of a complication from the domain of physics: 'this is more complicated due to this tube bending'. Thus the structure of the exposition belies the complex referent the speaker has, and the listener has little opportunity to make the meaning that the speaker intends.

This is an issue that deserves more attention, as it is seen that physicists, as other natural and engineering scientists, reify their mathematical formalisms in terms of the physics phenomena they are describing. And vice versa, in the physics phenomena they are dealing with they see mathematical formalisms. But here, in terms of the structure of exposition and with a pedagogical research interest, we can point out that expositions based on such dual conceptualisations give a taken-for-granted character to the listener's referent, and thus detracts from sense-making potential.

### *Making talk and making sense*

We have sought to relate our data and analysis to an understanding of language of relevance for this pedagogical research. There is a good deal of writing on language and its relation to thought from the socio-cultural movement. In line with these, we wish, at this stage of the argument, at least, to

> keep thought and communication (which is a social act) apart, not to see them as reflections of one another. One reason for this is that it is often difficult to achieve an explicit explanation of something that one understands or knows about. We can often use terms and concepts correctly and [only] long afterwards be able to explain them in a precise way. […] From a sociocultural perspective it is more reasonable to see communication as a situated act that is dynamic and partly unpredictable. We negotiate about – in Vygotsky's terminology – sense and meaning with our discussion partners. (Säljö, 2000, p 116, our translation)

This strengthens our characterisation of both the interview and the interviewer's role in it, and also giving priority to the structure of the talk rather then trying to infer its meaning. But it does not help us in analysis of the sorts of talk.

Mercer (2000), also writing from a socio-cultural perspective, makes a classification of types of normal talk, which might, one could think, be relevant for our classification of types of 'physics talk', being typifications of 'similar types of conversation'. He classifies conversation into cumulative, exploratory and disputational forms. In *cumulative* talk, 'speakers build on each other's contributions, add information of their own and in a mutually supportive, uncritical way construct share knowledge and understanding' (p30). *Exploratory* talk is 'that in which partners engage critically but constructively with each other's ideas. Relevant information is offered for joint consideration. Proposals may be challenged and counter-challenged, but if so reasons are given and alternatives offered. Agreement is sought as a basis for joint progress, Knowledge is made publicly accountable and reasoning is visible in the talk' (p 98). *Disputational* talk is characterised 'by an unwillingness to take on the other person's point of view, and the consistent reassertion of one's own.. […] It makes joint activity into a competition rather than a co-operative endeavour'.

When we look at our discussions of physics we see that large tracts that are different from any of these three types of talk. The initial exchanges and the closing-off might well be of a cumulative nature, exchanging information. Disputational talk is avoided, at least in the cases we have, which are typical for discussions between teachers and students or between colleagues and the cultural expectation is to maintain equanimity. Exploratory talk comes closest to the discussions we have seen, with the exception that rather than holding a common goal, the two parties probably have partially different goals: to expound and to understand. Proposals are certainly questioned and counter-questioned, but not with the purpose of making *common joint* progress: rather the two parties are on different paths of understanding.

The sort of talk we see here is, then, somehow special for the scientific discourse, and qualitatively different from everyday talk. It is aimed at making sense of a complex and abstract phenomenon, both for the person who holds the stage and the person who is paying attention. We can suppose that in his or her own way the speaker is picking up on aspects of this phenomenon and expressing them, or languaging them, in one way or another – and these ways constitute the outcome space. The listener is picking up on the same selection of words, the same ordering, the same sketches and formal mathematics, even the same structuring of pauses and gestures; but not necessarily making the same sense of them. Hence the listener's (generally the interviewer's, here) comments and questions. Making sense is a joint quest, even though the sense made is unlikely to be identical.

### *Languaging objects of knowledge, constituting knowledge objects*

Now we wish to move away from the structure of the data we have analysed, primarily from physicists at different stages in their careers speaking of a physics problem or topic that is close to their research potential, towards the meaning such talk of physics can have for the listener. The students are asked to engage in discussion of quite a basic but widely applicable physics phenomenon and the researchers are asked to talk of their own current research. These topics are, for the interviewees, 'objects of knowledge', in the term used by Svensson and Anderberg (in press), which they are trying to impart to a willing listener. An object of knowledge, here, is that to which one's cognitive awareness is directed, the phenomenon one is thinking of, or reading about, or speaking of – here the problem given to the students or the research topic chosen by the researchers; the discussion is a case of 'expressing and understanding objects of knowledge through language' (Svensson & Anderberg, in press). While a good deal of research has focused on the ways in which students make sense of, or understand, or comprehend the objects of knowledge that are described in texts, this, now, becomes a study of the ways in which such objects of knowledge are voiced by people in what can be construed as a pedagogical situation.

We see three pedagogical situations that commonly occur in the working life of the physicist that parallel the interview/discussion that provides our data and analysis.

1. The senior-junior research discussion, as in supervision of new research students or assistants
2. The oral examination of senior undergraduates by their teachers or tutors
3. Exposition and discussion in lectures and seminars

In each of these, one party is expected to speak of an object of knowledge for the benefit or critical consideration of the other. We can relate these pedagogical situations to the notion of 'a knowledge object', introduced by Entwistle and Marton as 'a tightly integrated and structured set of interconnecting ideas and data which together make up our personal understanding' (Entwistle and Marton, 1994). It is 'constructed by the individual learner out of a idiosyncratic combination of academic knowledge and personal experience in the process of making sense of the material being learned' (Entwistle, Marton & Entwistle, 1993). In their work it refers primarily to the multi-faceted, quasi-sensuous object that is the result of intense study by university students in the run-up to their final examinations. Booth and Ingerman (in press) have extended the use of the term to the extensive body of knowledge that is built up over a year of study in a physics programme, and they analyse the nature of the variation of such knowledge objects. Here we would extend its meaning yet again, to denote a relatively short-term and intensive consideration of an orally presented object of knowledge.

A 'knowledge object' is not to be confused with an 'object of knowledge'. In both cases the term '[is not] intended to suggest something that can be passed directly from the teacher to learner' (Entwistle, Marton & Entwistle, 1993): a knowledge object is an experiential and constituted understanding, and an object of knowledge is an abstracted and reified understanding. While the interviewee in this study is being asked to voice an object of knowledge, then, the interviewer is constructing his own knowledge object, based on the interconnecting ideas and data he finds there and his own personal experience. In an analogous pedagogical situation, it is the teacher who presents the object of knowledge and the student who constitutes a knowledge object.

Let us look at the forms of exposition and their potential for supporting understanding of the object of knowledge, or in these new terms, constituting a knowledge object.

Category A, Expounding in bits, as such is pretty hopeless! No-one listening to such a description of the problem or research topic would be able to make sense of it; the knowledge object would be no better than a clump of unrelated bits . Of course, in an overall exposition, a period of seeking a reasonable starting point for an unexpected question or problem can well have this form, and leads soon to a more coherent exposition of another form.

Expositions of the form of Category B, Expounding in a single perspective, can provide the listener with an extensive and/or intensive picture of a particular aspect of the object of knowledge, and affords the constitution of a knowledge object comprising just that aspect. As pointed out earlier, the speaker can all the time have a hidden private referent to the problem or research in another domain, which cannot be spied by the listener; nor can it be integrated into the emergent knowledge object

Expositions that fall into Category C, Expounding with multiple perspectives, in contrast, exposes the listener to more than one perspective on the object: physics and mathematics, or theory and experiment, for example. The speaker relates features of the problem or research seen in one perspective to relevant features of the other: how a particular feature of a physics phenomenon is represented in a developing mathematical formalism, for example, or how a mathematical solution relates to the structure of the research problem in hand. The listener can broaden his or her view of the problem or research, or begin to fill in the ties between aspects, and a knowledge object of a multiple-sided and integrated character is afforded.

Expositions that fall into Category D, Expounding through contextualisation, again offer the listener more than one view of the problem or research, now not only directly related to the object but also related to new contexts, or possibly to further examples that bear some relevance. Again, the listener has the opportunity to relate aspects of the problem or research topic to one another, now through finding ties in related objects or different contexts or the personal experience of the alternative contexts spoken of, and the constitution of a knowledge object that stands out against its various contexts is afforded.

Now the categories of exposition can be seen as a hierarchy of affordance of the constitution of an integrated knowledge object. This argument is in line with the variation theory of learning (Marton & Trigwell, 2000), where the teacher offers greater or lesser variation in perspectives on the phenomenon she is teaching, and thus affords more or less opportunity for seeing the phenomenon in a meaningful way.

If we return to our analogous pedagogical situations, we can see the opportunity to offer advice. The expounder should, at least, be aware of the virtues and shortcomings of his or her form of exposition, related to the affordance for learning thus provided. It is hard to make a positive case for the exposition of type category A, but the other three have their virtues. However, to support the constitution of nuanced, integrated knowledge objects, expositions of type category D have clear strengths over those of type category C, which have strengths over those of type category B. In contrast, for support of extended or intensified understanding of a delimited feature of an object of knowledge in a single perspective, then exposition of type category B is superlative.

Problematisation, consideration, and variation is the sum of the advice we can offer.

## *Conclusion*

The study, analysis, results and discussion we have presented here represent our insights into an important aspect of becoming a physicist – talking of physics in order for someone else to make sense of it. The study deserves to be extended to an analysis of exposition in other, naturalistic situations, such as in lectures, in seminars, in doctoral defences, and in conference presentations, which it is our intention to proceed with.

## *Acknowledgements*

First and foremost we want to thank the physicists and physics students who gave their time and attention to this investigation. Acknowledgement is also due to the Chalmers University of Technology Foundation, for financing the first author (Ingerman) and to the Knowledge Foundation of Sweden who financed the second author while working on this. The Centre for Educational Development at Chalmers is thanked for a grant to cover expenses. We are most grateful to Ference Marton for his particularly insightful comments on the draft of the paper, and to Cedric Linder, Anders Berglund and Tom Adawi for discussions at various stages of the study.